\documentclass[11pt]{report}
\usepackage{color}
\usepackage{graphicx}
\usepackage{harvard}
\pdfoutput=1

\def\farcs{\hbox{$.\!\!^{\prime\prime}$}}

  \font\sixrm=cmr6

\begin{document}

\baselineskip=.60cm

\centerline{ \Large{\bf Asteroid 2007 WD$_5$ will not impact Mars on
January 30! }}

\vspace{0.4cm}

\centerline{by}

\vspace{0.4cm}

\centerline{Ma{\l}gorzata Kr{\'o}likowska and Grzegorz Sitarski}

\vspace{0.4cm}

\centerline{Space Research Centre, Polish Academy of Sciences,}
\centerline{Bartycka 18A, 00-716 Warsaw, Poland}

\centerline{e-mail: mkr@cbk.waw.pl; sitarski@cbk.waw.pl}

\vspace{0.4cm}

{\small \centerline{A B S T R A C T}

\vspace{0.4cm}

The Monte Carlo method of the nominal orbit clonning was applied to
the case of 2007~WD$_5$, the asteroid from the Apollo group.
Calculations based on 33 observations from the time interval of
2007\,11\,08~-–~2008\,01\,02 showed that the asteroid will pass near
planet Mars at the minimum distance of $10.9\pm 2.9$~R$_{\rm Mars}$,
what implies that probability that 2007~WD$_5$ strike the planet
decreased to the value of 0.03\% from the value of about 3--4\%
previously announced by NASA. The additional observations taken on
January~3--9 reduce further the asteroid's impact chances,
effectively to nil: the asteroid will pass near planet Mars at the
minimum distance of $8.4\pm 1.1$~R$_{\rm Mars}$.

}

\vspace{0.4cm}

\centerline{\bf 1. Introduction}

\vspace{0.4cm}

The asteroid, known as 2007\,WD$_5$, was discovered on November 20,
2007 by Boattini  of the NASA-funded Catalina Sky Survey on Mount
Lemmon, near Tucson (Arizona, USA) using a 1.5\,m telescope. The
size of asteroid was estimated as $\sim $50~\,m.

Next day, Chesley and Chodas (2007) from the NASA/JPL Near-Earth
Object Program Office announced that newly discovered asteroid which
passed close to the Earth in November, will pass very close to Mars
in late January, and there is a chance that it could hit that
planet.


On December 28, 2007, Yeomans et al. (2007) informed on Web page
that three pre-discovery observations of asteroid, taken on November
8, 2007 were located by Puckett in the archive of the Sloan Digital
Sky Survey II at the Apache Point Observatory. They refined the
orbit of 2007\,WD$_5$ and found that the probability of Mars impact
on January 30, 2008 increased to 4\%, whereas the uncertainty region
of the asteroid position during the Mars encounter has shrunk to
400\,000 km along a very narrow ellipsoid. They stressed that the
uncertainty region of the asteroid position during the Mars
encounter could shrink more -- to the region which will no longer
intersect with the planet surface.

The last (at the moment of writing this text) report from NASA/JPL
\cite{Ches08} -- based on the additional observations of the
asteroid between Dec. 29 and Jan. 9 -- significantly diminished the
asteroid chances to strike Mars to approximately 0.01\%. Chesley et
al.~\citeyear{Ches08} estimated that asteroid 2007~WD$_5$ most
likely will pass about 8~Mars radii, r$_{\rm Mars}=$8\,R$_{\rm
Mars}$, from the planet's center on Jan. 30 and with 99.7\%
confidence, the asteroid should be no closer than 1.2\,R$_{\rm
Mars}$ from the planet's surface.

According to NeoDys (2008) calculations based on the observation arc
from 2007\,11\,08 to 2008\,01\,09 the asteroid will pass near Mars
at the nominal distance of $2.204\cdot 10^{-4}$\,AU (9.7\,R$_{\rm
Mars}$), whereas the minimum possible distance is $8.815\cdot
10^{-5}$\,AU (3.9\,R$_{\rm Mars}$). These estimates suggest that
asteroid 2007~WD$_5$ will pass Mars even in the greater distance
than was suggested by Chesley et al.~\citeyear{Ches08}.

The aim of this note is to present how our method of the nominal
orbit clonning allows to determine the set of impact orbits as
well as the probability of this impact for the case of 2007~WD$_5$.

\vspace{0.4cm}

\centerline{\bf 2. The method}

\vspace{0.4cm}

The present investigations are based on the archive positional
observations taken from the NeoDys (Near Earth Object - Dynamic
Site, University of Pisa, Italy) publicly available on the Web at
\newline http://newton.dm.unipi.it/neodys/mpcobs/2007WD5.rwo. The whole
observational material contains 46 observations covering the time
period from November~8,~2007 to January~9, 2008.

The equations of asteroids motion have been integrated numerically
using the recurrent power series method (Sitarski \citeyear{Sit89},
\citeyear{Sit02}), taking into account the perturbations by all the
nine planets and by the Moon. The planetary coordinates were taken
from the Warsaw numerical ephemeris DE405/WAW of the Solar System,
consistent with a high accuracy with the JPL ephemeris DE405
\cite{Sit02}.

We derived the nominal orbits using the least squares orbit
determination method for four different sets of residuals given in
the second column of Table~\ref{tab:cases}. The respective solutions
are denoted as A, B, C, and D, where solutions B and C differ only
in a number of residuals taken into account for the orbit fitting:
three residuals of significantly larger values in comparison to the
rest of data were excluded in the case~C. The nominal rms's are
given in column~4 of Table~\ref{tab:cases}.

The sample of 10\,000 clones of each nominal orbit was selected
randomly  according to the normal distribution in the 6D-space of
orbital elements (the method is described in details by Sitarski
\citeyear{Sit98}). In the column~5 of Table~\ref{tab:cases}  upper
limits for the rms of the true (but unknown) orbit are given for the
confidence level of 99\%.
\newline The orbit clones from the samples A--D were integrated forwards up to
the close encounter with Mars on January 30.

\vspace{0.4cm}

\centerline{\bf 2. Results}

\vspace{0.4cm}

\begin{figure}
\begin{center}
\includegraphics[width=15.0cm]{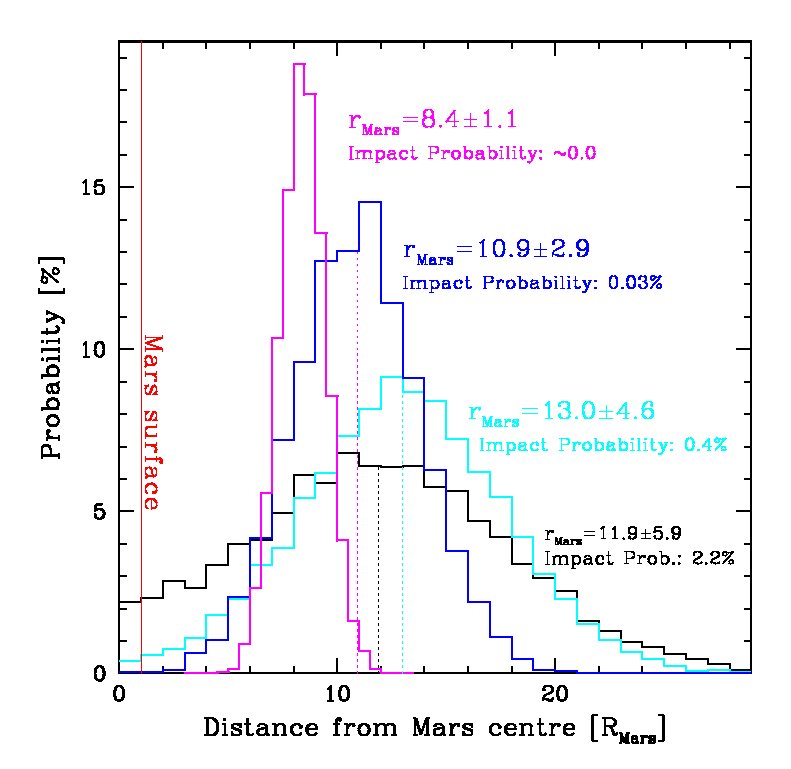}
\end{center}
\caption{Distribution of the minimum distance of the asteroid
2007~WD$_5$ from the centre of Mars in 2008 01 30.5 derived for the
samples of 10\,000 clonned orbits. Black, cyan, blue and magenta
distributions correspond to the solutions~A, B, C and D,
respectively.} \label{fig:MarsDist}
\end{figure}

We obtained 220 impact orbits for the Mars encounter on January 30,
2008 by numerical integrations of 10\,000 orbital clones in case~A.
The distribution of minimum asteroid distance from Mars are
presented for the solution~A by black histogram in
Fig.~\ref{fig:MarsDist}. Thus, the impact probability of $2.2\cdot
10^{-2}$ was calculated for this arc of observations. However, the
refinement of the asteroid's orbit using also the observations taken
on January 2 significantly reduces the chance of impact on Mars.
Cyan, blue and magenta histograms in Fig.~\ref{fig:MarsDist}
represent the distribution of minimum asteroid distance from  Mars
for solutions~B, C and D, respectively. One can see, that for the
solution~C the impact probability decreases to the value of $3\cdot
10^{-4}$ (we found three impact orbits among the sample of 10\,000
clones), and for the case~D -- the impact seems to be impossible.

\begin{figure}
\begin{center}
\includegraphics[width=10.5cm]{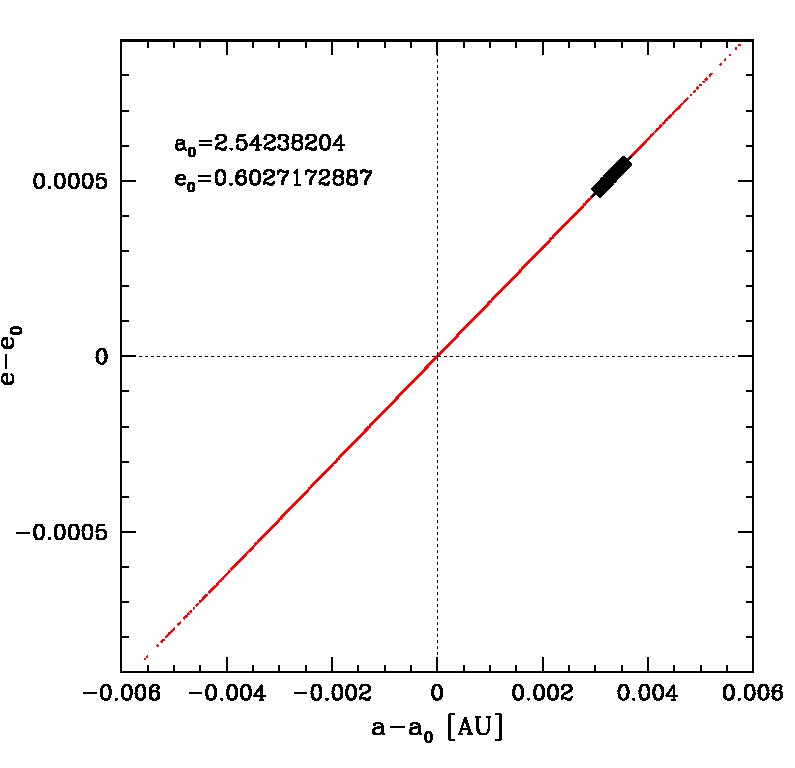}
\includegraphics[width=10.5cm]{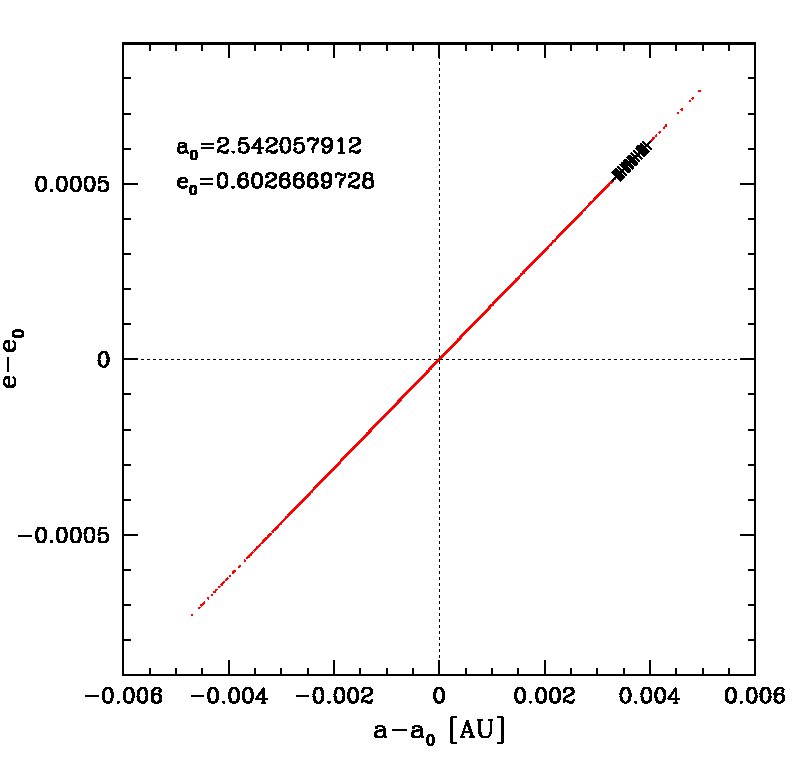}
\end{center}
\caption{ Projection onto the a-e plane in the 6-dimensional space
of possible osculating orbits of 2007~WD$_5$ obtained for solution A
(upper panel) and solution B (lower panel) Sample of 10\,000 clonned
orbits is given by red points, while the impact orbits are given as
black crosses. The plot is centered on the values of semimajor axis,
$a_0$,  and eccentricity, $e_0$, of the respective nominal orbits
(epoch: 2007 10 27). } \label{figa:ae}
\end{figure}

\begin{figure}
\begin{center}
\includegraphics[width=10.5cm]{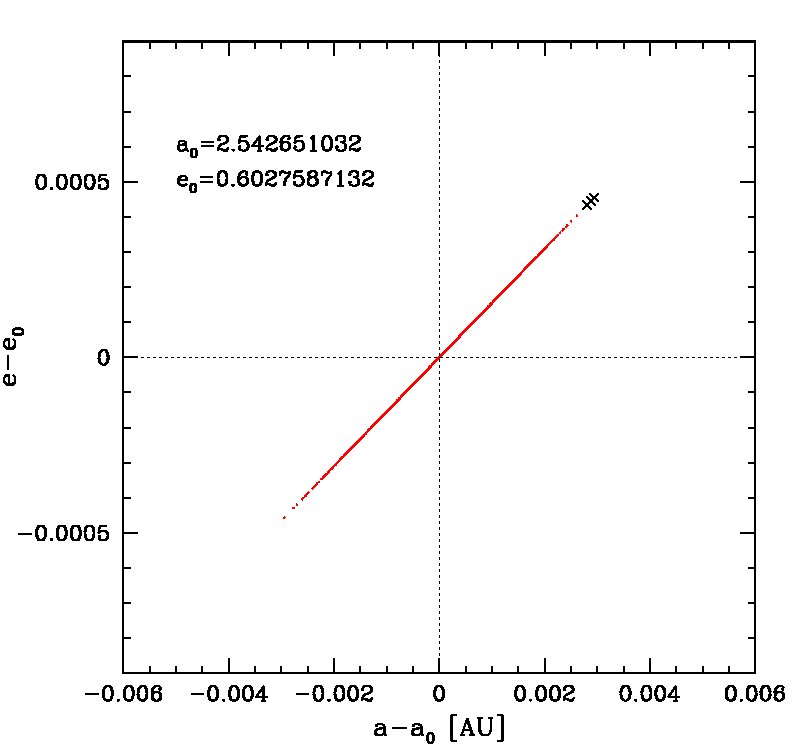}
\includegraphics[width=10.5cm]{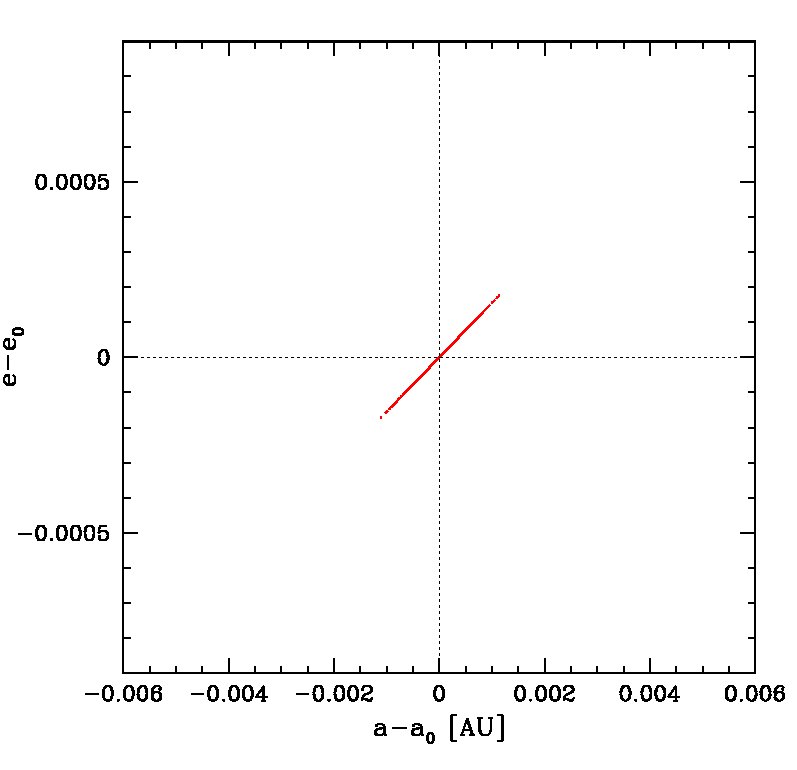}
\end{center}
\caption{Projection onto the a-e plane in the 6-dimensional space of
possible osculating orbits of 2007~WD$_5$ obtained for solution C
(upper panel) and solution D (lower panel). Sample of 10\,000
clonned orbits is given by red points, while the impact orbits are
given as black crosses. The plots are centered on the values of
semimajor axis, $a_0$,  and eccentricity, $e_0$, of the respective
nominal orbits (epoch: 2007 10 27).} \label{figb:ae}
\end{figure}

Gaussian distribution of r$_{\rm Mars}$ and the impact probability
are given in the columns 6 and 7 of Table~1 and in
Fig.~\ref{fig:MarsDist}. Projections of the 6D ellipsoid of orbital
elements onto the a-e plane for the solutions A--D are shown by red
points in Fig.~\ref{figa:ae}--~\ref{figb:ae}. The shrinking of the
uncertainty ellipsoid with the refinement of 2007~WD$_5$ nominal
orbit and its moving away from the impact region are clearly
visible. For the longest arc of observations (case~D) the asteroid
will pass near planet Mars at the minimum distance of $8.4\pm
1.1$~R$_{\rm Mars}$ (magenta distribution in Fig~\ref{fig:MarsDist})
what implies that with 99.7\% confidence, the asteroid should be no
closer than 4.1\,R$_{\rm Mars}$ from the planet's surface.

\vspace{0.1cm}

Thus, our calculations indicate that the possibility of asteroid
2007~WD$_5$ collision with Mars seems to be ruled out.

\vspace{0.3cm}

\begin{table}
\caption{{Description of four orbital solutions for 2007~WD$_5$ and
the derived minimum distance for asteroid close encounter with Mars
on January 30, 2008. To estimate the value of rms$_{\rm true}$
(column~5) the confidence level of 99\% was assumed.}}
\label{tab:cases} \vspace{0.10cm} {\setlength{\tabcolsep}{1.0mm} {
\sixrm {\small
\begin{center}
\begin{tabular}{ccccccc} \hline \hline
 Solution &  Observational  & Number  & Mean   & rms$_{\rm true}$ & Minimum                & Impact      \\
          &   interval      & of      & res.   & not greater      & distance               & probability \\
          &                 & res.    &        & than             & [R$_{\rm Mars}$]       &  [\%]       \\
\hline
 A & 2007\,11\,08 -- 2007\,12\,31 & 64 & 0\farcs 521 & 0\farcs 585  & $11.9\pm 5.9$          & 2.2          \\
 B & 2007\,11\,08 -- 2008\,01\,02 & 68 & 0\farcs 512 & 0\farcs 572  & $13.0\pm 4.6$          & 0.4          \\
 C & 2007\,11\,08 -- 2008\,01\,02 & 65 & 0\farcs 308 & 0\farcs 346  & $10.9\pm 2.9$          & $\sim$ 0.03  \\
 D & 2007\,11\,20 -- 2008\,01\,09 & 89 & 0\farcs 268 & 0\farcs 293  & ~$8.4\pm 1.1$          & $\sim$ 0.0   \\
 \hline
\end{tabular}
\end{center}
}
}}
\end{table}

\vspace{0.5cm}

\vspace{0.4cm}

\centerline{\bf Acknowledgements}

\vspace{0.4cm}
We are grateful to Dr Ireneusz W\l odarczyk from Astronomical
Observatory of the Chorz\'ow Planetarium for very valuable
discussion. This work was partly supported by the Polish Committee
for Scientific Research (the KBN grant 4~T12E~039~28).

\end{document}